\documentclass[12pt,a4paper]{iopart}
\usepackage{graphicx,cite}
\usepackage{bm,bbm}% bold math
\usepackage{iopams}

\begin{document}
% ===== \newcommand ==================================================
% vectors
\renewcommand{\vec}[1]{\boldsymbol{#1}}
\newcommand{\qq}{\vec{q}}
\newcommand{\pp}{\vec{p}}
\newcommand{\abs}[1]{\left| #1 \right|}
\newcommand{\ee}{\mathrm{e}}
\newcommand{\dd}{\mathrm{d}}
\newcommand{\ii}{\mathrm{i}}
\renewcommand{\vec}[1]{\boldsymbol{#1}}
\newcommand{\rr}{\vec{r}}
\newcommand{\ff}{\vec{f}}
\newcommand{\generator}{\vec{g}}
\newcommand{\xx}{\vec{x}}
\newcommand{\yy}{\vec{y}}
\newcommand{\zz}{\vec{z}}
\newcommand{\mm}{\vec{m}}
\newcommand{\RR}{\vec{R}}
\newcommand{\eps}{\varepsilon}
\newcommand{\p}{\partial}
\newcommand{\ad}[1]{\mathrm{ad}_{#1}}
\newcommand{\order}[1]{\mathcal{O}\!\left({#1}\right)}
\newcommand{\Lie}[2]{\left[#1,#2\right]}
\newcommand{\Dop}[1]{\mathcal{D}_{\!#1}}
\newcommand{\Lop}[1]{\mathcal{L}_{\!#1}}
\newcommand{\strich}[1]{\left.#1\right|}
\newcommand{\Htilde}{\tilde{H}}
\newcommand{\Hhat}{\hat{H}}
\newcommand{\Hdensity}{\mathcal{H}}
\newcommand{\AZ}[1]{``#1''}
\newcommand{\Ng}{{N_{\!g}}}
\newcommand{\nmax}{{n_\mathrm{max}}}

\title[TST for wave packet dynamics.\\ II. Thermal decay of BECs with long-range interaction]{Transition state theory for wave packet dynamics.\\ II. Thermal decay of Bose-Einstein condensates with long-range interaction}

\author{Andrej Junginger, Markus Dorwarth, J\"org Main, and G\"unter Wunner}

\address{1. Institut f\"{u}r Theoretische Physik, Universit\"{a}t Stuttgart, 70550 Stuttgart, Germany}

\date{\today}


% Abstract %%%%%%%%%%%%%%%%%%%%%%%%%%%%%%%%%%%%%%%%%%%%%%%%%%%%%%%%%%%% 
\begin{abstract}
We apply transition state theory to coupled Gaussian wave packets and calculate thermal decay rates of Bose-Einstein condensates with additional long-range interaction. The ground state of such a condensate is metastable if the contact interaction is attractive and a sufficient thermal excitation may lead to its collapse. The use of transition state theory is made possible by describing the condensate within a variational framework and locally mapping the variational parameters to classical phase space as has been demonstrated in the preceding paper [A. Junginger, J. Main, and G. Wunner, preceding paper, submitted to J. Phys. A]. We apply this procedure to Gaussian wave packets and present results for condensates with monopolar $1/r$-interaction comparing decay rates obtained by using different numbers of coupled Gaussian trial wave functions as well as different normal form orders.
\end{abstract}

\maketitle

% Introduction %%%%%%%%%%%%%%%%%%%%%%%%%%%%%%%%%%%%%%%%%%%%%%%%%%%%%%%%
\section{Introduction}
Since their first experimental realization in 1995 \cite{Anderson1995} Bose-Einstein condensates (BECs) have become an active field of theoretical and experimental investigations. Moreover, BECs with additional long-range interaction are of special interest, because the interactions can be tuned from predominantly short-range to the dominance of the long-range interaction by manipulating the $s$--wave scattering length via Feshbach resonances. The latter allow varying the contact interaction in strength as well as in sign. In case of a negative scattering length, i.e. an attractive interaction, the ground state of the BEC is metastable so that the condensate may decay by collapsing after a sufficient thermal excitation.

The thermal decay rates of BECs without long-range interaction have already been estimated by Huepe \etal\ \cite{Huepe1999,Huepe2003} within a simple variational ansatz of a single Gaussian wave function. This approach can, of course, also be applied to condensates with long-range interaction, but it will, because of its simplicity, only yield qualitative results.

In this paper, we present an improvement by using an extended variational ansatz with \emph{coupled} Gaussian trial wave functions, which are described in the framework of a time-dependent variational principle and which have proven their capability to reproduce the numerically exact results or even to exceed them \cite{Rau2010a, Rau2010b}. Within this variational ansatz, the BEC exhibits two stationary states, one of which corresponds to its metastable ground-state, while the other one is an excited state of saddle-centre-\ldots-centre type. The decay rate of the condensate can, thus, be calculated by means of transition state theory (TST) \cite{Waalkens2008} because the collapsing BEC has to cross this saddle in the subspace of the variational parameters and the decay rate is given by the flux over the saddle. 

Although classical TST requires the knowledge of a Hamilton function $H(\qq,\pp)$ given in phase space coordinates $\qq,\pp$, its application is made possible by locally mapping the variational parameters to action variables of the classical phase space in the vicinity of the fixed points. This local mapping to phase space is performed with the use of a normal form expansion of the equations of motion determining the time evolution of the variational parameters as well as the respective mean-field energy functional. 

For systems with known classical Hamiltonian this procedure has been shown to reproduce the decay rates of the classical and the quantum normal forms in the limits of narrow and broad wave functions, respectively \cite{Junginger2011b}, and, moreover, it well applies to systems where such a Hamilton function is not directly accessible as it is the case for the variational approach to BECs with coupled Gaussian wave functions.

In order to demonstrate the applicability to BECs and to calculate their decay rates, our paper is organized as follows: First, we review the description of the variational ansatz in the framework of a time-dependent variational principle as well as the procedure of locally mapping the variational parameters to phase space. Then, we illustrate the calculation of the thermal decay rates by applying TST and at the end present and discuss the results for  BECs with monopolar $1/r$-interaction.

\section{Theory}
We consider a condensate consisting of $N$ bosons which exhibit an additional long-range $1/r$-interaction as has been proposed by O'Dell \etal\ \cite{ODell2000}. Such systems have not yet been experimentally realized but because of the spherical symmetry of this interaction, they form an important model system.

At ultra-low temperatures this quantum gas can be described by a single wave function $\psi(\rr,t)$ whose time evolution is given by the extended Gross-Pitaevskii equation (GPE)
\begin{equation}
 	\hat{H} \psi(\rr,t) = \ii \p_t \psi (\rr,t), 
	\label{eq-GPE}
\end{equation} 
where the mean-field Hamiltonian 
\begin{equation}
	\hat{H} = -\Delta + V_\mathrm{t} + V_\mathrm{c} + V_\mathrm{m}
	\label{eq-Hamilton-operator}
\end{equation}
describes the interaction with an external trapping potential $V_\mathrm{t}$, the inter-atomic contact interaction $V_\mathrm{c}$ with the $s$-wave scattering length $a_\mathrm{sc}$, and the long-range monopolar interaction $V_\mathrm{m}$:
\begin{eqnarray}
	V_\mathrm{t}	&=	N^4\gamma^2 \rr^2,					\\
	V_\mathrm{c}	&=	8\pi a_\mathrm{sc} \abs{\psi(\rr)}^2,		\\
	V_\mathrm{m}	&=	-2 \int \! \dd^3 \rr' \frac{\abs{\psi(\rr')}^2}{\abs{\rr-\rr'}}.
	\label{eq-potentials}%
\end{eqnarray}

To obtain the GPE in this dimensionless form, we introduce \AZ{atomic} units \cite{Cartarius2008a} with the help of the constant $u$ which determines the strength of the attractive monopolar inter-atomic interaction $V(\rr,\rr')=-u/\abs{\rr-\rr'}$ \cite{ODell2000}. Lengths are measured by means of the \AZ{Bohr radius} $a_u=\hbar^2/(mu)$, energies in units of the \AZ{Rydberg energy} $E_u=\hbar^2/(2ma_u^2)$, and times in units of $t_u=\hbar/E_u$. In addition, we apply a particle number scaling according to
\begin{eqnarray}
	\rr 			&\to	N^{-1} a_u \rr	,			\\
	a_\mathrm{sc}	&\to N^{-2} a_u a_\mathrm{sc},	\\
	t			&\to N^{-2} t_u t,				\\
	E			&\to N^2 E_u E	,				\\
	\psi 		&\to (a_u N)^{-3/2}\psi,
	\label{eq-particle-number-scaling}%
\end{eqnarray}
which eliminates the explicit occurrence of the particle number $N$ in the GPE.

\subsection{Application of transition state theory to the Gross-Pitaevskii equation}

The usual way of solving the GPE, equation \eref{eq-GPE}, is either by performing an imaginary time evolution on a grid or by integrating it outward with initial values for the wave function and its derivative \cite{Papadopoulos2007}. The former method can also be applied to more general geometries and more complicated interaction potentials such as the dipole-dipole interaction, while the latter is limited to the case of effectively one-dimensional systems like the radially symmetrical BEC investigated in this paper.

In order to apply TST it is furthermore essential to precisely define the transition state of the system and therefore to find the unstable stationary solution of the GPE. However, since an imaginary time evolution will only yield the ground state of the system this method cannot be applied. Thus, the approach may be limited to the simplest case of a monopolar BEC in a radially symmetrical trap.

If the stable $(s)$ and the unstable $(u)$ solution of the GPE are found the corresponding local normal form Hamiltonians are required in addition. Their lowest-order quadratic approximations will take the form \cite{Waalkens2008}
\begin{eqnarray}
	H^{(s)}(\qq,\pp) &= H^{(s)}_0 + \sum_{j=1}^d  \ii \omega^{(s)}_j p_jq_j, 
	\label{eq-Hamiltonians-quadratic-a}\\
	H^{(u)}(\qq,\pp) &= H^{(u)}_0 + \lambda p_1q_1 + \sum_{j=2}^d  \ii\omega^{(u)}_j p_jq_j,
	\label{eq-Hamiltonians-quadratic-b}%
\end{eqnarray}
respectively, where $\lambda$ is the eigenvalue corresponding to the decay channel of the BEC. Assuming their knowledge, the evaluation of the respective phase space integrals \cite{Toller1985} is trivial and one obtains the decay rate
\begin{equation}
	\Gamma = \frac{\omega_1^{(s)}}{2\pi} \ee^{-\beta \left(E^{(u)}_0-E^{(s)}_0 \right)} \prod_{j=2}^d \frac{\omega_j^{(s)}}{\omega_j^{(u)}},
        \label{eq-gamma-quadratic}
\end{equation}
which is a generalization of the formula given in reference \cite{Huepe2003} to $d$ degrees of freedom. 

The frequencies $\omega_j^{(s,u)}$ can, in principle, be obtained by the Bogoliubov-de Gennes (BdG) equations. However, this would present difficulties since they exhibit an unbounded spectrum so that the decay rate given by equation \eref{eq-gamma-quadratic} will, in general, either diverge or vanish. Moreover it is questionable, to what extent a local quadratic approximation will yield appropriate results after integration over the whole phase space. 

An extension of the Hamiltonians \eref{eq-Hamiltonians-quadratic-a} and \eref{eq-Hamiltonians-quadratic-b} to higher-order terms would require the treatment of small perturbations to the solution of the GPE in higher-order approximations than the linear one resulting in the BdG equations. In conclusion, a numerically exact approach for the application of TST to the GPE does not appear promising.

All these problems can, however, be circumvented if one treats the GPE within a variational framework: This is, in principle, not limited by any restrictions concerning the geometry of the system and the number of degrees of freedom, respectively. Moreover, it can be applied to various interaction potentials, allows a rather simple determination of the stable and unstable stationary solutions of the GPE and with it a precise definition of the transition state. Even the problem of unbounded spectra does not occur because one makes use of a finite set of variational parameters. Furthermore, an extension to higher-order normal form Hamiltonians can be carried out since the respective terms are related to numerically easily computable higher derivatives of the equations of motion which determine the time evolution of the variational parameters. 

\subsection{Variational approach to the Gross-Pitaevskii equation}

In the following, we will describe the condensate by means of a variational ansatz
\begin{equation}
	\psi(\rr,t) = \psi(\rr,\zz(t)) = \sum_{i=1}^\Ng \exp(A_i r^2 + \gamma_i)
	\label{eq-wave-function-parameter}
\end{equation}
in the form of a radially symmetrical Gaussian wave packet. Here, the complex and time-dependent variational parameters $A_i$ and $\gamma_i$ determine the width, the phase and the weight of each Gaussian, and we summarize all these parameters in the vector $\zz=(\zz_1, \ldots,\zz_\Ng)^T$ with $\zz_i=(A_i,\gamma_i)$.

An approximate solution of the time-dependent GPE \eref{eq-GPE} within the parameter subspace of the wave function \eref{eq-wave-function-parameter} is given by the McLachlan variational principle \cite{McLachlan1964}
\begin{equation}
	I = \| \ii \phi - \Hhat \psi \| \stackrel{!}{=} \mathrm{min.}
	\label{eq-McLachlan-variational-principle}
\end{equation}
where the quantity $I$ is minimized with respect to $\phi$ and $\phi=\dot{\psi}$ is set afterwards. Its application to the parametrized wave function \eref{eq-wave-function-parameter} yields the set of first-order differential equations \cite{Fabcic2008}
\begin{equation}
	K \dot{\zz} = - \ii \vec{h},
	\label{eq-zdot}
\end{equation}
which determines the time evolution of the variational parameters, and where the matrix $K$ and the vector $\vec{h}$ are defined by
\begin{eqnarray}
	K_{mn} &= \int \! \dd^3\rr~ \left(\frac{\p \psi}{\p z_m}\right)^* \frac{\p \psi}{\p z_n} , \label{eq-def-K}\\
	{h}_m  &= \int \! \dd^3\rr~ \left(\frac{\p \psi}{\p z_m}\right)^* \hat{H}  \psi .
	\label{eq-def-h}%
\end{eqnarray}%
The mean-field energy of the condensate is given by the expectation value
\begin{equation}
	E(\zz) = \int \! \dd^3 \rr ~  \psi^*(\rr)  \left( -\Delta + V_\mathrm{t} + \frac{1}{2} (V_\mathrm{c} + V_\mathrm{m}) \right) \, \psi(\rr)	
	\label{eq-meanfield-energy}
\end{equation}
and is also a function of the variational parameters.

For the ansatz with coupled Gaussian wave functions \eref{eq-wave-function-parameter}, all integrals occurring in equations \eref{eq-def-K}--\eref{eq-meanfield-energy} can be calculated analytically. However, since these calculations have been illustrated in detail elsewhere and are not subject of this paper, we refer the reader to references \cite{Rau2010a, Rau2010b} for their evaluation.

\subsection{Mapping to phase space}
\label{sec-mapping-to-phase-space}

The application of TST in phase space \cite{Waalkens2008} requires knowledge of a (local) Hamilton function which describes the dynamics of the BEC in the vicinity of the unstable fixed point corresponding to the \AZ{activated complex}. Such a Hamiltonian can easily be obtained even globally if one uses a \emph{single} Gaussian to approximate the BEC's wave function \cite{Cartarius2008a}. In contrast to that, this is not possible in the case of \emph{coupled} wave functions, as we use them in this paper. We therefore apply the procedure presented in reference \cite{Junginger2011b} to construct a \emph{local} Hamilton function in the vicinity of the fixed points and in this section give a short overview on the steps performed (see reference \cite{Junginger2011b} for details).

To obtain the local Hamilton function, which, equivalently to equations \eref{eq-zdot} and \eref{eq-meanfield-energy}, describes the dynamics and the energy of the system, one first Taylor expands the equations of motion \eref{eq-zdot} in the vicinity of a fixed point $\zz_0$ up to the order $\nmax$ and splits the expansion into its real and imaginary part. This yields the real vector field
\begin{equation}
	\dot{\xx} = \vec{a}(\xx) = \sum_{n=1}^\nmax \vec{a}_n (\xx)
	\label{eq-DGL-series}
\end{equation}
with $\xx=(\mathrm{Re}(z_1-z_{01}), \mathrm{Im}(z_1-z_{01}), \ldots, \mathrm{Re}(z_d-z_{0d}), \mathrm{Im}(z_d-z_{0d}))$ being the deviation of the variational parameters from the fixed point and $\vec{a}_n (\xx)$ summarizing all terms homogeneous of degree $n$. 

In the next step, equation \eref{eq-DGL-series} is diagonalized with respect to its linear part $\vec{a}_1(\xx) = A_1 \cdot \xx$ and to further \AZ{simplify} the higher-order terms, a near-identity transformation $\xx\to\yy$ given by (cf. reference \cite{Murdock2010})
\begin{equation}
	\xx = \vec{\phi}_\eps(\yy), \qquad \xx = \vec{\phi}_{\eps=0}(\yy) = \yy
	\label{eq-change-of-coordinates}
\end{equation}
is performed. Here $\xx = \vec{\phi}_\eps(\yy)$, which gives the identity transformation for $\eps=0$, is a solution of the differential equation ${\dd \xx}/{\dd \eps} = \vec{g} (\xx)$ and for an appropriate choice of the generating function $\generator(\xx)$ brings equation \eref{eq-DGL-series} into the form ($i=1,\ldots,d$)
\begin{eqnarray}
	\dot{x}_{2i-1} &= 	\sum_{\mm}	\beta_{\mm (2i-1)}	x_{2i-1}^{m_{2i-1}} x_{2i}^{m_{2i}-1} 	\prod_{j\neq i} (x_{2j-1} x_{2j})^{m_{2j}},	\label{eq-structure-monomials-x-a}\\
	\dot{x}_{2i} &=  	\sum_{\mm} 	\beta_{\mm (2i)~~~} x_{2i-1}^{m_{2i-1}-1} x_{2i}^{m_{2i}} 	\prod_{j\neq i} (x_{2j-1} x_{2j})^{m_{2j}}.
	\label{eq-structure-monomials-x-b}
\end{eqnarray}
This form is due to the fact that the eigenvalues of equation \eref{eq-zdot} occur pairwise with different sign and is obtained as long as the eigenvalues are in rational independence and do not fulfil the condition of \AZ{resonance}
\begin{equation}
	\vec{\lambda} \vec{m} - \lambda_i = 0
	\label{eq-resonance-condition}
\end{equation}
for integer vectors $\mm$, with $\abs{\mm}\leq n_\mathrm{max}$.

Analogously, the energy functional \eref{eq-meanfield-energy} is also Taylor expanded and transformed according to the change of variables in the equations of motion which results in the expansion
\begin{equation}
	E = \sum_{\mm} \xi_{\mm} \prod_{j} (\tilde{q}_j \tilde{p}_j)^{m_j}.
     \label{eq-transformed-energy-functional}
\end{equation}
after introducing canonical coordinates $ \tilde{q}_i = x_{2i-1}$ and momenta $\tilde{p}_i = x_{2i}$. Moreover, with the latter definition the equations of motion \eref{eq-structure-monomials-x-a}--\eref{eq-structure-monomials-x-b} can easily be integrated to a common Hamilton function
\begin{equation}
	\Htilde = \sum_{\mm} \frac{\beta_{\mm (2i)}}{m_{2i}} (\tilde{q}_i \tilde{p}_i)^{m_{2i}} \prod_{j\neq i} (\tilde{q}_j \tilde{p}_j)^{m_{2j}}
	\label{eq-Htilde}
\end{equation}
according to Hamilton's equations if the coefficients $\beta_{\mm}$ satisfy the conditions of integrability
\begin{eqnarray}
	\beta_{\mm (2i-1)} &= - \beta_{\mm (2i)}, 						\\
	\frac{\beta_{\mm (2i-1)}}{m_{2i}} &= \frac{\beta_{\mm (2i'-1)}}{m_{2i'}},	\\
	\frac{\beta_{\mm (2i)}}{m_{2i-1}} &= \frac{\beta_{\mm (2i')}}{m_{2i'-1}}
	\label{eq-integration-conditions}%
\end{eqnarray}
for all $i,i'=1,\ldots,d$ ($i\neq i'$). In order to guarantee both, the satisfaction of these conditions of integrability as well as the equivalence of the integrated Hamiltonian $\Htilde$ with the energy functional \eref{eq-transformed-energy-functional} an additional transformation is necessary. For this purpose, we scale the phase space variables with time-independent functions $\nu_{q_i}(\qq,\pp)$ and $\nu_{p_i}(\qq,\pp)$ according to
\begin{equation}
	\tilde{q}_i = \nu_{q_i}(\qq,\pp)\, q_i~,	\qquad
	\tilde{p}_i = \nu_{p_i}(\qq,\pp)\, p_i.
	\label{eq-nu-scaling}
\end{equation}
with the constraint of their product
\begin{equation}
	\mu_i (\qq,\pp) = \nu_{q_i}(\qq,\pp) \, \nu_{p_i}(\qq,\pp)  = 1 + \sum_{\mm} \mu_{\mm} \prod_j (q_j p_j)^{m_j}
	\label{eq-mu-power-series}
\end{equation}
to be a formal power series of the products $q_j p_j$, as well as an appropriate choice of the $\mu_{\mm}$ \cite{Junginger2011b} finally guarantees the equivalence of the integrated Hamiltonian \eref{eq-Htilde} with the transformed energy functional:
\begin{equation}
	H \left(\vec{J} \right) = E\left(\vec{J}\right) = \Htilde \left(\vec{J}\right)
	\label{eq-Hamiltonian}
\end{equation}
Here, action variables  $J_i={q}_i {p}_i$ and ${J}_i= \ii {q}_i {p}_i$, respectively, have been introduced depending on whether the corresponding eigenvalue of the linearised equations of motion is real or purely imaginary. Equation \eref{eq-Hamiltonian} finally serves as classical Hamilton function in the sense that it locally reproduces the energy of the system and its Hamiltonian equations of motion describe the dynamics in the vicinity of the fixed point equivalently to equation \eref{eq-zdot}.

\subsection{Thermal decay rates}

Within the variational approach to monopolar BECs using coupled Gaussian wave functions, the set of differential equations \eref{eq-zdot} exhibits two fixed points \cite{Rau2010b}. One of them is stable corresponding to the metastable ground state of the condensate and the other one is of saddle-centre-\ldots-centre type corresponding to an unstable excited state.

\begin{figure}[t]
	\centering	
	\includegraphics[width=.7\columnwidth]{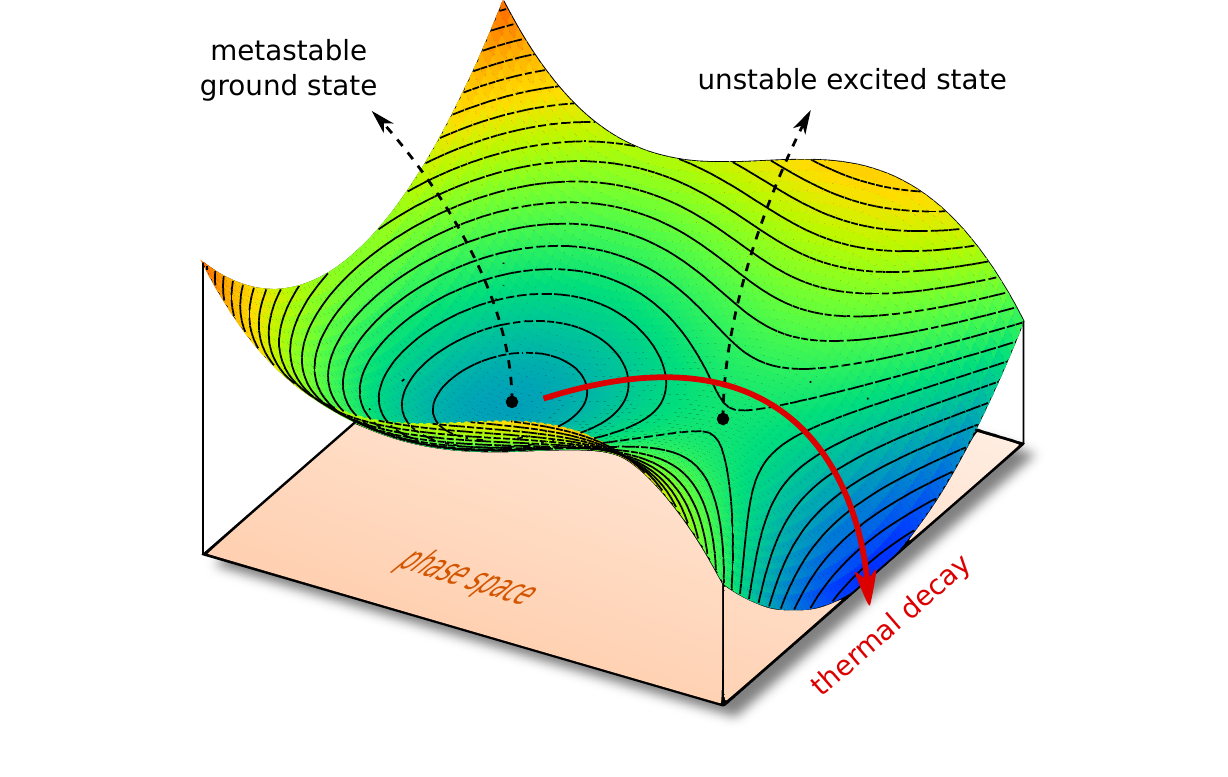}
	\caption{Schematic drawing of the phase space structure of the constructed Hamilton function in equation \eref{eq-Hamiltonian}. The metastable ground state of the BEC corresponds to a local minimum, and classical decay is possible after thermal excitation. If the only decay channel (solid arrow) requires crossing a saddle point in phase space the thermal decay rate is given by the Boltzmann average of the flux over this saddle.}
	\label{fig-zerfallsbild}
\end{figure}

These properties, of course, also hold after having applied the near-identity transformation described above in order to locally map the variational parameters to phase space. The constructed Hamilton function in phase space, thus, takes the form depicted in figure \ref{fig-zerfallsbild}, featuring a local minimum and a saddle. The latter has precisely one unstable direction and can, therefore, be used to divide the phase space into a region of \AZ{reactants} formed by the metastable BEC and a region of \AZ{products} in the form of the collapsed condensate. Calculating the decay rate is, thus, possible by applying TST (see reference \cite{Waalkens2008}) in association with the constructed Hamiltonian, because the only possibility of the BEC to collapse is by crossing this saddle, and the decay rate is given by the flux over it.

% Figure: Phase space structure %%%%%%%%%%%%%%%%%%%%%%%%%%%%%%%%%%%%%%%%%%%%%%%%

At a fixed energy, the directional flux through the dividing surface between \AZ{reactants} and \AZ{products} is given by \cite{Waalkens2008, Waalkens2004, MacKay1990}
\begin{equation}
	f(E) = (2\pi)^{d-1} \mathcal{V}(E)
	\label{eq-flux-over-saddle}
\end{equation}
with $\mathcal{V}(E)$ being the phase space volume of the actions $(J_2,\ldots,J_d)$ which is enclosed by the contour $H(0,J_2, \ldots, J_d) \leq E$ and $J_1=0$ corresponding to the \AZ{unstable direction} of the saddle. If the condensate is in contact to a bath of finite temperature the thermal decay rate is then given by the Boltzmann average of equation \eref{eq-flux-over-saddle}. After a short calculation, this yields (cf. reference \cite{Toller1985})
\begin{equation}
	\Gamma = \frac{1}{2\pi \hbar ^d \beta Z_0} \int  \ee^{-\beta H(0,J_2,\ldots,J_d)} \, \dd J_2 \ldots \dd J_d,
	\label{eq-decay-rate-1}
\end{equation}
where $\beta=1/k_\mathrm{B}T$, and $Z_0$ is the canonical partition function. Because nearly all states will be localized in the vicinity of the ground state, we can well approximate the latter by 
\begin{equation}
	Z_0 = \frac{1}{\hbar^d} \int \dd J'_1\, \ldots \dd J'_d ~ e^{-\beta H'(J'_1,\ldots,J'_d)}
	\label{eq-partition-function}
\end{equation}
with $H'(J'_1,\ldots,J'_d)$ being the normal form expansion at the local minimum. Furthermore taking into account the particle number scaling \eref{eq-particle-number-scaling}, the thermal decay rate is given by 
\begin{equation}
	\Gamma = \frac{1}{2\pi \beta } \frac{\int  \ee^{-N^2\beta H(0,J_2,\ldots,J_d)} \dd J_2 \ldots \dd J_d }{\int \dd J'_1\, \ldots \dd J'_d ~ e^{-N^2\beta H'(J'_1,\ldots,J'_d)} },
	\label{eq-thermal-decay-rate}
\end{equation}
where we identify $N^2\beta$ as the particle number scaled inverse temperature.

However, both integrals in equation \eref{eq-thermal-decay-rate} will, in general, not converge, which is due to the fact that the normal form expansion has been truncated at the order $\nmax$. We will therefore restrict the area of integration to the condition 
\begin{equation}
	\omega_i = \frac{\p H}{\p J_i} \geq 0 
	\label{eq-condition-frequencies}
\end{equation}
for all $i$, in view of the fact that all frequencies occurring on the tori in phase space have to be non-negative.

\section{Results and discussion}
\label{sec-results}

In order to calculate the thermal decay rate of BECs with monopolar interaction, we first determine the stable and the unstable fixed point of the equations of motion, equation \eref{eq-zdot}, for given physical parameters $N^4\gamma^2$ and $a_\mathrm{sc}$ and Taylor expand these in the vicinity of the fixed points up to a chosen order $\nmax$. Then, we proceed as described in section \ref{sec-mapping-to-phase-space} to map the variational parameters to classical phase space variables and obtain the corresponding local Hamilton functions $H(\vec{J})$ and $H'(\vec{J}')$, respectively, from the transformed mean-field energy functional. With their knowledge, the decay rate is calculated from equation \eref{eq-thermal-decay-rate} under the constraint \eref{eq-condition-frequencies}.

As has been shown by Rau \etal\ \cite{Rau2010b} the main contribution from the extended variational ansatz occurs when the number of wave functions is increased from $\Ng=1$ to $\Ng=2$. We will therefore restrict ourselves to that case in the following and, moreover, to Hamiltonians up to fourth order of the action-variables where we observe convergence.

For $\Ng=2$ coupled Gaussians we have eight real variational parameters $\xx$, one of which is fixed by normalizing the wave function to $\int\!\dd^3\rr\,\abs{\psi(\rr,t)}^2=1$ and another corresponds to a global phase that can be set to zero, so that we are left with six independent ones. Determining the corresponding classical Hamiltonian in fourth order approximation in $\vec{J}$ is already non-trivial since, in this case, the expansion of the mean-field energy functional up to eighth order in the variational parameters $\xx$ (scalar valued polynomial with 3003 terms) and that of the equations of motions in seventh order of $\xx$ (vector valued polynomial with 10\,290 terms) are required. After the mapping to phase space, these are simplified to a fourth order polynomial of $\vec{J}$ with 35 terms, which is a reduction of the number of monomials by altogether 99.74\%.

% figure: comparison between 1 and 2 Gaussians %%%%%%%%%%%%%%%%%%%%%%%%%%%%%%%%%%%%%%%%%%
\begin{figure}[t]
	\centering
	\includegraphics[width=.7\columnwidth]{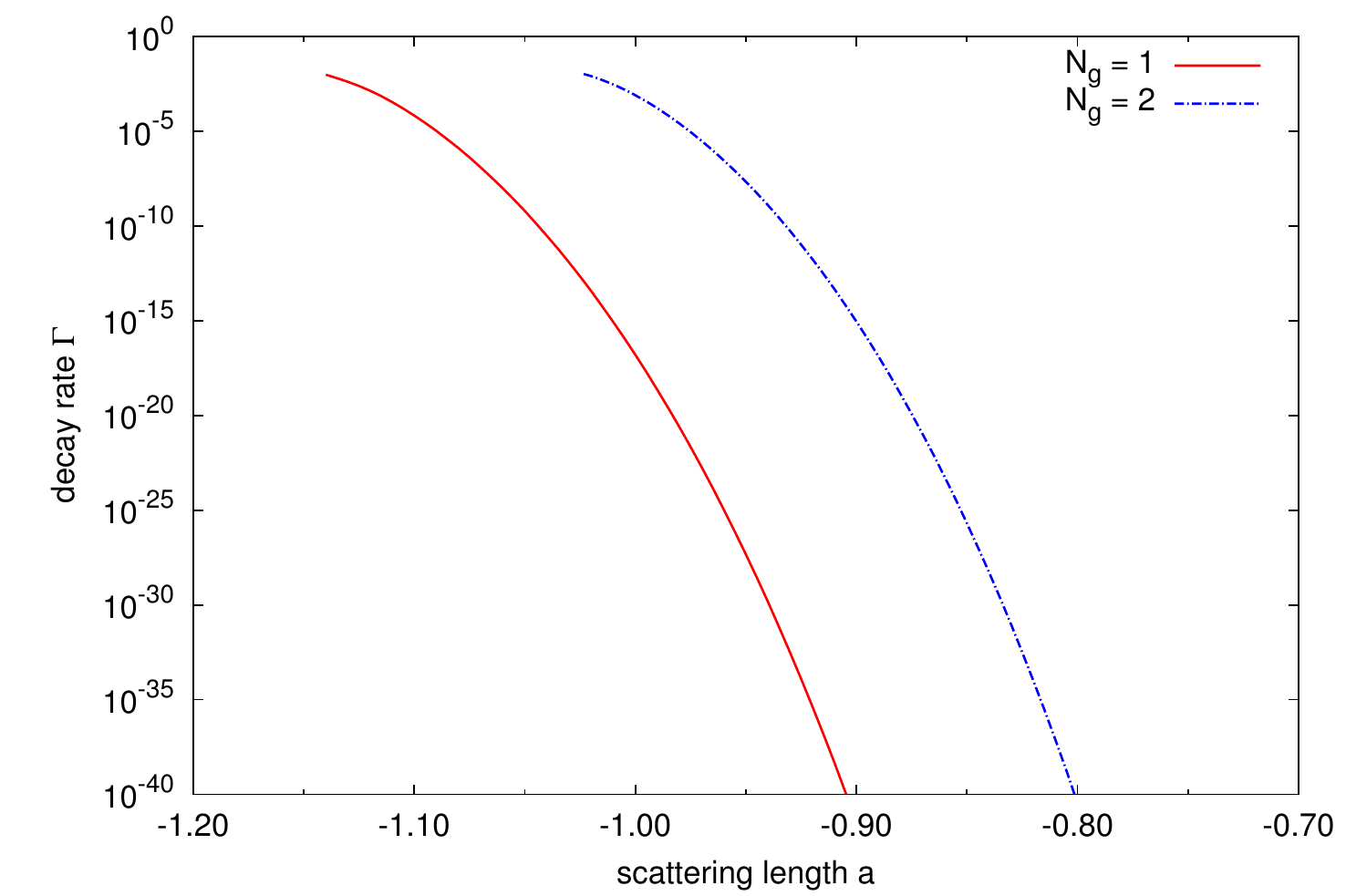}
	\caption{Comparison of the thermal decay rate of a monopolar BEC described with $\Ng=1$ (solid line) and $\Ng=2$ (dashed-dotted line) Gaussian wave functions in dependence of the scattering length $a_\mathrm{sc}$ in third order normal form of the action variable. The data have been calculated for a particle number scaled inverse temperature of $N^2\beta=900$ and a trap frequency of $N^4\gamma^2=1.0\times 10^{-3}$. It can be seen that the critical scattering length is shifted to higher values when increasing the number of Gaussians and that, for a fixed scattering length, the decay rate rises by several orders of magnitude.}
	\label{fig-comparison}
\end{figure}

Note that a monopolar BEC, as investigated in this paper, features the phenomenon of self-trapping under certain conditions \cite{Papadopoulos2007}, i.e. an external trap is not necessary to keep the condensate stable. However, at least a weak trap is required here to avoid a dissolving of the BEC, which otherwise would be a second decay channel. An external trap, thus, guarantees that the only decay mechanism of the BEC is its collapse.

Figure \ref{fig-comparison} shows the thermal decay rate of a monopolar BEC calculated in third order normal form in $\vec{J}$ using a single Gaussian trial wave function (solid line) and $\Ng=2$ coupled Gaussians (dashed-dotted line). For the calculation, we used a particle number scaled inverse temperature of $N^2\beta=900$ and a weak trap with a frequency of $N^4\gamma^2=1.0\times 10^{-3}$. 

One consequence of the use of coupled Gaussian wave functions is that the critical scattering length $a_\mathrm{crit}$ below which the condensate cannot exist is shifted to larger values (cf. reference \cite{Rau2010b}). For the parameters used here, this is the case from $a_\mathrm{crit} \approx -1.145$ for a single Gaussian trial wave function to  $a_\mathrm{crit} \approx -1.024$ for the two coupled ones. Figure \ref{fig-comparison} reveals this behaviour for the whole curve and shows that the thermal decay rate calculated with a single Gaussian, as it has also been used in references \cite{Huepe1999,Huepe2003} for BECs without long-range interaction, underestimates the result of the extended variational ansatz by several orders of magnitude for a fixed value of the scattering length $a_\mathrm{sc}$. Considering the point of the critical scattering length, the decay rate changes only very little compared to a single Gaussian, and the general dependence of the decay rate on the scattering length is retained exhibiting a rapid monotonic increase when decreasing the scattering length. This increase, however, becomes weaker when one approaches the critical value.

% figure: different normal form orders %%%%%%%%%%%%%%%%%%%%%%%%%%%%%%%%%%%%%%%%%%%%%%%%%%
\begin{figure}[t]
	\centering
	\includegraphics[width=.7\columnwidth]{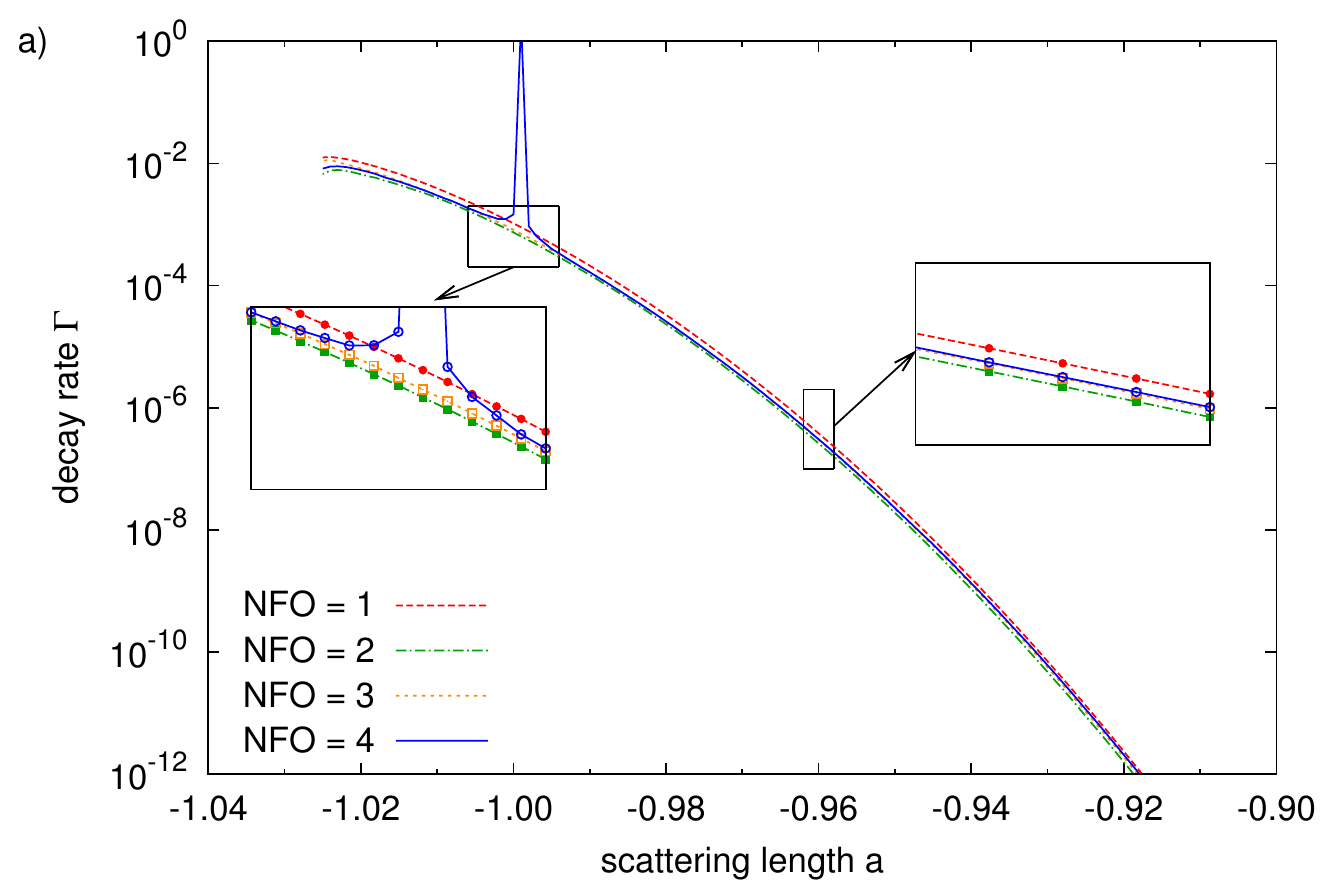}
	\includegraphics[width=.7\columnwidth]{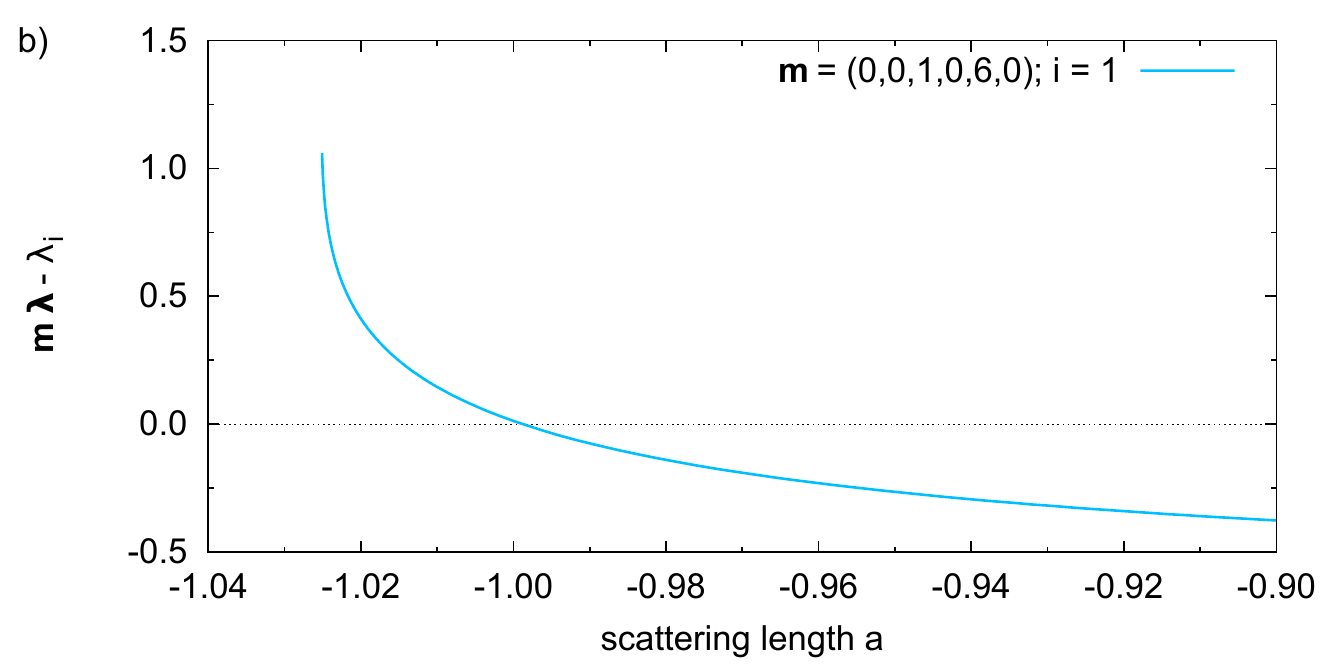}
	\caption{a) Thermal decay rate of a monopolar BEC described with $\Ng=2$ coupled Gaussian wave functions in dependence of the scattering length $a_\mathrm{sc}$ and normal form orders (NFO) $1$ to $4$ of the action variables. Temperature and trap frequency are the same as in figure \ref{fig-comparison}. Right inset: The thermal decay rates obtained from the third- and fourth-order normal form Hamiltonian cannot be distinguished any more, indicating convergence. Left inset: At $a_\mathrm{sc}\approx -0.999$ the eigenvalues are close to \AZ{resonance} and the normal form expansion as well as the decay rate diverge. As shown in b) this is the case because the condition of resonance \eref{eq-resonance-condition} is numerically fulfilled for $\mm=(0,0,1,0,6,0)$ for this set of physical parameters.}
	\label{fig-orders}
\end{figure}

Figure \ref{fig-orders}a shows the thermal decay rates for different normal form orders. The calculations have been performed for $\Ng=2$ coupled Gaussians and for the same physical parameters used in figure \ref{fig-comparison}. The first-order approximation (dashed line) overestimates the decay rate over the whole range of the scattering length, whereas using the normal form Hamiltonian in second order in $\vec{J}$ (dashed-dotted line), we observe the smallest values throughout. However, the results calculated by the third- and fourth-order Hamiltonian (dotted and solid lines) cannot be distinguished within the line width of the plot (right inset in figure \ref{fig-orders}a), indicating convergence.

At a scattering length of about $a_\mathrm{sc}\approx -0.999$, one observes a strong deviation of the calculated decay rate in the fourth order approximation from all the other curves (left inset in figure \ref{fig-orders}a), which is in contrast to the behaviour all along the rest of the investigated range of the scattering length. As shown in figure \ref{fig-orders}b, the eigenvalues $\lambda_i$ of the linearised equations of motion which are used for the normal form expansion run into \AZ{resonance}, i.e. equation \eref{eq-resonance-condition} is fulfilled within the numerical accuracy for the integer vector $\mm=(0,0,1,0,6,0)$ in seventh order of the variational parameters, $\abs{\mm}=7$ (corresponding to the fourth order in $\vec{J}$ after integration). This leads to the divergence of the fourth order normal form Hamiltonian and with it the decay rate at $a_\mathrm{sc}\approx -0.999$.

% figure: different temperatures %%%%%%%%%%%%%%%%%%%%%%%%%%%%%%%%%%%%%%%%%%%%%%%%%%%%%%%
\begin{figure}[t]
	\centering
	\includegraphics[width=.7\columnwidth]{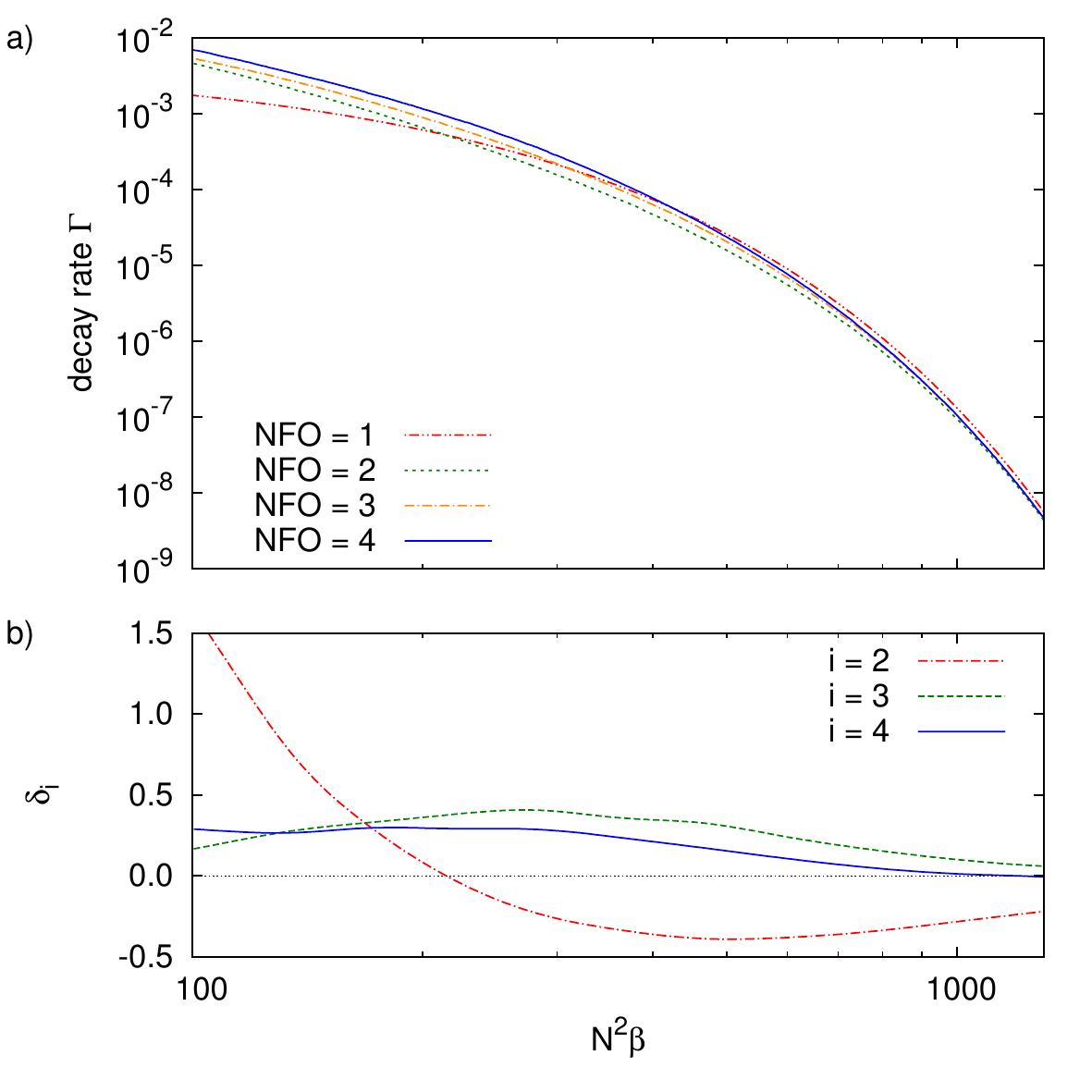}
	\caption{a) Thermal decay rate of a monopolar BEC for a fixed scattering length of $a_\mathrm{sc}\approx -0.96$, a trap frequency of $N^4\gamma^2=1.0\times 10^{-3}$ and normal form orders $1$ to $4$ in dependence of the particle number scaled temperature $N^2\beta$. The BEC is described with $\Ng=2$ coupled Gaussian wave functions. b) The relative deviation $\delta_i$ defined in equation \eref{eq-relative-deviation} is used in order to estimate the convergence of the procedure.}
	\label{fig-temp}
\end{figure}

Moreover, the convergence behaviour of the decay rate with increasing normal form order strongly depends on the temperature of the system (see figure \ref{fig-temp}a). While, we observe fast converging results for low temperatures and large particle numbers, respectively, i.e. large values $N^2\beta$, where the decay rates calculated from the third- and fourth-order normal forms match within the line width for $N^2\beta \gtrsim 800$ the convergence becomes worse when decreasing the scaled inverse temperature $N^2\beta$. For $N^2\beta \lesssim 200$ the calculations even show a monotonic increase of the decay rate with higher normal form order.

In order to estimate the convergence of our results, we use the relative deviation
\begin{equation}
	\delta_i = (\Gamma_{\mathrm{NFO}=i} -\Gamma_{\mathrm{NFO}=i-1} )/\Gamma_{\mathrm{NFO}=i-1}
	\label{eq-relative-deviation}
\end{equation}
shown in figure \ref{fig-temp}b. The corrections to the decay rate obtained from the second- ($i=2$) and the third- ($i=3$) order normal form are significant throughout, while this is only true for low $N^2\beta$ for the fourth order ($i=4$). For large values of $N^2\beta$ the corrections quickly shrink and in case of $N^2\beta \gtrsim 1000$ these are of the relative order of $10^{-4}$ to $10^{-5}$, clear evidence again of the convergence of the decay rate in fourth-order approximation in the low-temperature regime.

\section{Summary and outlook}

We have demonstrated the applicability of a variational approach to classical transition state theory by means of calculating thermal decay rates of Bose-Einstein condensates with an additional long-range interaction using \emph{coupled} Gaussian wave functions. This procedure has proven as a powerful tool for that purpose: For the extended variational ansatz with coupled Gaussians, we observed convergence in eighth order of the variational parameters at cold temperatures and high particle numbers, respectively. Moreover, the results show that previous estimations using single Gaussian wave functions \cite{Huepe1999, Huepe2003}, on the one hand, reveal a good qualitative agreement with those obtained from the extended ansatz but, on the other, underestimate the thermal decay rate by several orders of magnitude for a fixed value of the scattering length.

To further improve the results and to also achieve convergence for higher temperatures the procedure can be extended to the use of more than two coupled Gaussian wave functions and higher normal form orders. This is also necessary in order to compute decay rates of experimentally accessible dipolar BECs \cite{Griesmaier2005} where the symmetry breaking dipole-dipole interaction and the occurrence of blood-cell shaped condensates \cite{Ronen2007} require up to six coupled and non-radially symmetrical Gaussians to reach the accuracy of the numerical results \cite{Rau2010b}. Convergence of the decay rate is, finally, expected when increasing both the number of Gaussians as well as the normal form order.

\section*{Acknowledgement}

This work was supported by Deutsche Forschungsgemeinschaft. A.\,J.\ is grateful for support from the Landesgraduiertenf\"orderung of the Land Baden-W\"urttemberg. We also thank Thomas Bartsch for fruitful discussions.

% Bibliography ==========================================
\section*{References}
% \bibliographystyle{unsrt}
% \bibliography{Literature.bib}

\end{document}